# Fibre architecture modification to improve the tensile properties of flax- reinforced composites

Rishad Rayyaan[a], William Richard Kennon[a], Prasad Potluri[a], Mahmudul Akonda[a]

[a] School of Materials, The University of Manchester, Oxford Road, Manchester, M13 9PL

## Abstract

As far as the tensile properties of natural fibres as reinforcements for composites are concerned, flax fibres will stay at the top-end. However, an efficient conversion of fibre properties into their corresponding composite properties has been a challenge, due to the fibre damages done through the conventional textile methods utilised to process flax. These techniques impart disadvantageous features onto fibres at both micro-, and meso-level, which degrade the mechanical performances of flax fibre reinforced composites (FFRC). Undulation of fibre is one of those detrimental features that occur during traditional fibre extraction and fabric manufacturing routes. The undulation or waviness causes micro-compressive defects or 'kink bands' in elementary flax fibres, which significantly undermines the performance of FFRC. Manufacturing flax fabric with minimal undulation could diminish the micro-compressive defects up to a substantial extent. In this research, nonwoven flax tapes of highly aligned flax fibres, blended with a small proportion of PLA (Polylactic Acid) have been manufactured deploying a novel technique. Composites reinforced from those nonwoven tapes have been compared with composites reinforced with woven Hopsack fabrics and warp knitted unidirectional (UD) fabrics from flax that are comprised of undulating fibres. The composites reinforced with the highly aligned tape have shown 49% higher fibre bundle strength, and 100% higher fibre bundle stiffness in comparison with that of the Hopsack fabric reinforced composites. The results have been discussed in the light of fibre undulation, elementary fibre individualisation, homogeneity of fibre distribution, extent of resin rich areas, and impregnation of the fibre lumens.

## 1 Introduction

Among the natural fibres that are currently being used as composites' reinforcements, flax falls into the top-end of the list, in terms of tensile strength and modulus [1]. However, composites reinforced with mineral fibres such as glass or carbon inevitably endow longitudinal tensile and compressive properties beyond the capability of composites reinforced with natural fibres, such as flax. In an effort to reduce this gap, several researchers have succeeded in improving the mechanical performance of flax fibre reinforced composites (FFRC) by modifying the flax fibre morphology through chemical treatment on fibres [2-6]. Chemical treatments, however, involve additional manufacturing steps, which increase manufacturing costs. Modification of fibre geometry can be an effective substitute method to improve the mechanical performance of flax fibres inside composites. The geometry of the fibres in composites such as fibre undulation directly influences the mechanical performance of the composites [7-19]. Currently, plain woven flax fabrics with undulating fibres are widely being used as composite reinforcements[1, 20]. The research discussed in this paper is focused on how to reduce fibre waviness for improvement of mechanical performances of FFRC. In pursuit to that, nonwoven flax tapes blended with a small proportion of PLA have been manufactured as reinforcements, using a custom-designed roller-drafting machine. There will also be discussion as to why reduction of waviness plays a vital role in performance enhancement of FFRC.

A strand of flax sliver is an assembly of technical fibres and a technical fibre in turn comprises from single number up to tens of elementary fibres[21]. Lengthwise technical fibres can be as long as the flax stem, and are bonded together with hemicellulose and pectin [21]. Inside technical fibres, these elementary fibres tend to overlap each other by a substantial length [21, 22]. The cross-section of an elementary fibre can have a pentagonal to octagonal shape, depending upon the cell-growth within the plant. An elementary fibre consists of an outer primary cell wall, an inner secondary cell wall (S2), and a hollow channel called the 'lumen', which runs through the core of the fibre. The secondary cell wall comprises micro-fibrils and accounts for the major portion of the elementary fibre's cross-section. The cellulosic micro-fibrils are crystalline and are spirally wound in a polysaccharide matrix of amorphous hemicellulose and lignin. The helical angle of the fibrils with respect to the fibre axis is +10° [23-25]. The arrangement of the micro-fibrils resembles a unidirectional composite structure, which confers a good tensile property of the fibres [24, 26-28]. Page et al. in 1977, showed that the elastic modulus of natural fibres decreases with the increase of the micro-fibrillar helix angle [29]. Hence, a low helix angle confers good strain to failure properties whilst minimising the compromise of tensile strength and longitudinal compressive strength.

When a flax fibre is bent by any mechanical mean, the cellulosic fibrils of the S2 wall become dislocated, and at the same point the hemicellulose that binds the micro-fibrils together fails. The lateral compressive force caused by bending does not result in failure of the micro-fibrils, rather they become slightly separated, and form cracks bridged by coarse fibrils. This structure looks like a crack and is termed a slip plane, or a node, or a kink band, as shown in Figure 8. Kink bands are micro-compressive defects of flax fibres [22, 27, 30, 31]. Bos (1999) showed that the gradual compressive force on the elementary fibre increases the number of kink bands [24].

Kink bands reduce the dry fibre's tensile strength as those defective areas potentially initiate failure under tensile loading. Inside the composites as well, the kinked areas are susceptible to damage initiation under tensile loading. For the case of dry flax fibres, if the fibre undergoes cyclic tensile loading, a reorientation in the micro-fibrils takes place (which are aligned at +10° to the fibre axis), and this results in strain-hardening by straightening out the kink bands. Eventually the non-linearity of the tensile curve of the dry fibres disappears and the modulus increases [25]. This 'strain hardening' does not occur within flax reinforced composites as the fibres are locked in the matrix. Therefore, upon tensile loading, stress-concentration develops around the kinked areas, which in turn initiates fibre-matrix de-bonding as well as micro-cracks within the matrix [25, 32-34].

Traditional textile processing for flax fibres such as breaking, scutching, hackling, drawing, spinning, and weaving involve a lot of fibre bending and fibre undulation, eventually resulting in developing kink bands. Also, fibre reinforcements inside a composite can withstand maximum load if aligned exactly with the loading direction. Therefore, a nonwoven flax tape with fibres parallel to each other is clearly a possible solution to nullify the effect of waviness up to a great extent. A number of researchers in recent times have conducted experiments on nonwoven tape [33, 35-39]. UD fabric with twistless yarn is another method to optimise fibres alignment, which has been investigated by Miao and Shan (2011) [38].

In this research, a highly aligned nonwoven tape, comingled with a small proportion of PLA has been manufactured with no noticeable out-of-plane waviness. A novel technique has been exploited here, and the compatibility of the newly made tape has been studied in comparison with plain woven fabric (Hopsack) and warp knitted unidirectional fabric (UD).

# 2 Materials and manufacturing

## 2.1 Fabrics

Four types of fabric have been used in this research namely Hopsack (plain woven Hopsack fabric), UD (warp knitted UD fabric made from twistless wrap-spun yarn, as shown in Figure 3), T180 (nonwoven tape of 180 mm width), and GVT (nonwoven tape attached with a glass fibre veil). Hopsack and UD were procured from a local company named 'Composite Evolution'. The other two structures were manufactured as a part of this research, using a novel technique. Table 1 shows the dry fabric specifications. Figure 1 shows the topology of the fabric surfaces whereas Figure 2 shows the cross-sections of all four kinds of fabric. The undulation angle shown by the warp yarns in the Hopsack dry fabric is 17±1° and inside the composite, the undulation angle was found to be 14±1°. Dry UD fabric shows an undulation angle averaging 6.1° whereas no discernible undulation can be found in the UD composites. The tapes also showed no noticeable undulation of the fibres, either in dry or in composite state.

Table 1   Dry fabric specification

| | Hopsack | UD | T180 | GVT |
|---|---|---|---|---|
| Yarn Linear density, in tex | 250 | 250 | n/a | n/a |
| Flax content, % | 88 | 84.53 | 90 | 89.5 |
| Glass content, % | n/a | n/a | n/a | 10.5 |
| Polyester content, % | 12 | 15.47 | n/a | n/a |
| PLA content, % | n/a | n/a | 10 | n/a |
| Fabric construction | 4×4 Hopsack | Warp knitted | Nonwoven tape | Nonwoven tape with surface veil |
| Areal density, in g/cm2 | 0.0519 | 0.0264 | 0.0158 | 0.049 |
| Fabric density, g/cc | 1.4845 | 1.4802 | 1.467 | 1.564 |
| Ends/inch | 24 | 24 | n/a | n/a |
| Pics/inch | 30 | 6 | n/a | n/a |

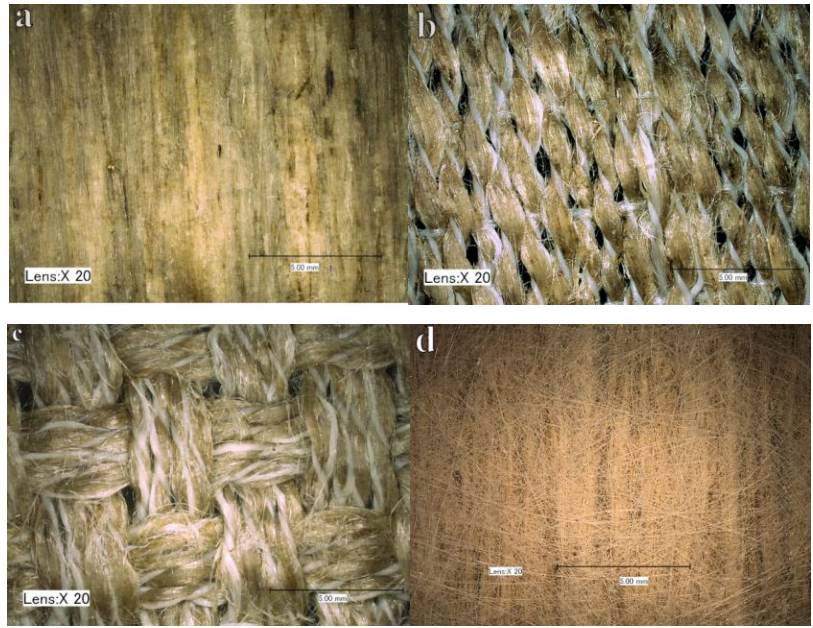

Figure 1   Optical images of flax fabrics (top view): (a) Nonwoven tape; (b) Warp knitted UD fabric; (c) Hopsack fabric; and (d) Nonwoven tape with surface veil.

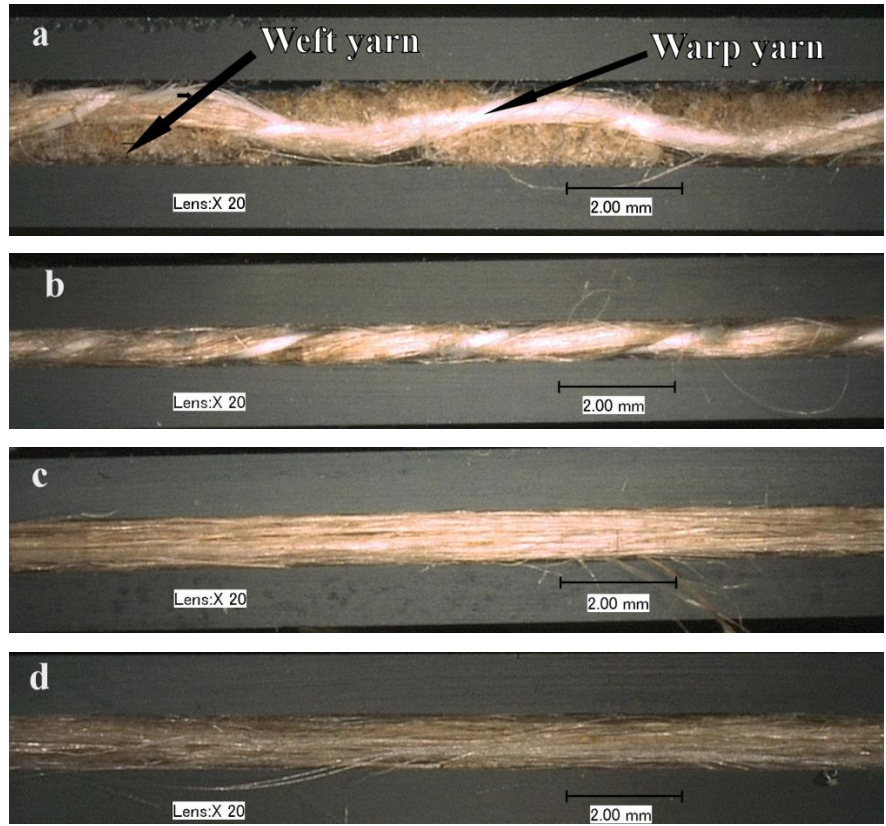

Figure 2  Cross-sectional images of flax fabrics captured by optical microscope: (a) Hopsack fabric; (b) UD; (c) T180; and (d) GVT.

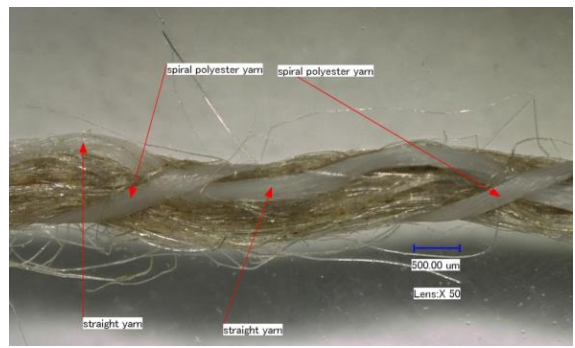

Figure 3  Twistless flax yarn.

## 2.2  Nonwoven tape manufacture

A commingled form of flax fibres with 10% PLA binder fibres (90:10 w/w), was produced using a specially designed draw-frame, attached with a calendaring machine, as depicted in Figure 4. Sequentially, this consisted of, what is termed, a 3-over-3 roller-drafting unit, a heating unit and then a pair of pressurised calendars. To begin the process of conversion, the blends are fed to the roller-drafting unit. The combined action of the paired rollers aligns the flax fibres in the machine direction (which is ultimately the axis of the resulting tape). This action is known as drafting, and each successive pair of rollers operates with a surface speed slightly faster than the proceeding pair, thereby attenuating the fibre mass as it passes from one roller pair to another. During drafting, the frictional contact between fibres induces the localised shear forces required to orientate the flax fibres uni-directionally. Drafting therefore simultaneously, thins and separates the flax/PLA blends into a fine sheet of fibres in addition to effecting fibre alignment. On leaving the drafting stage the thin sheet of fibres is heated just above the melting point of the PLA and additionally, is drafted by the faster surface speed of the calendar rollers, for further alignment of the flax fibres. The calendar rollers simultaneously apply a pressure of 3 bars to the flax sheet, melting the PLA fibres to the flax to produce a semi-consolidated tape. Partially consolidated tapes of 170 mm width, were produced at 170ºC. GVT was produced with glass surface veils on both sides of nonwoven tape and no PLA was blended with flax. To add surface veils, rolls of glass veils were mounted between the drafting and heating zone on both planes. Glass veils that were used contained a small proportion of adhesive materials, subjected to be activated during heating. These adhesive materials provided the required adhesive strength for the glass veils to be attached with the surfaces of the tapes.

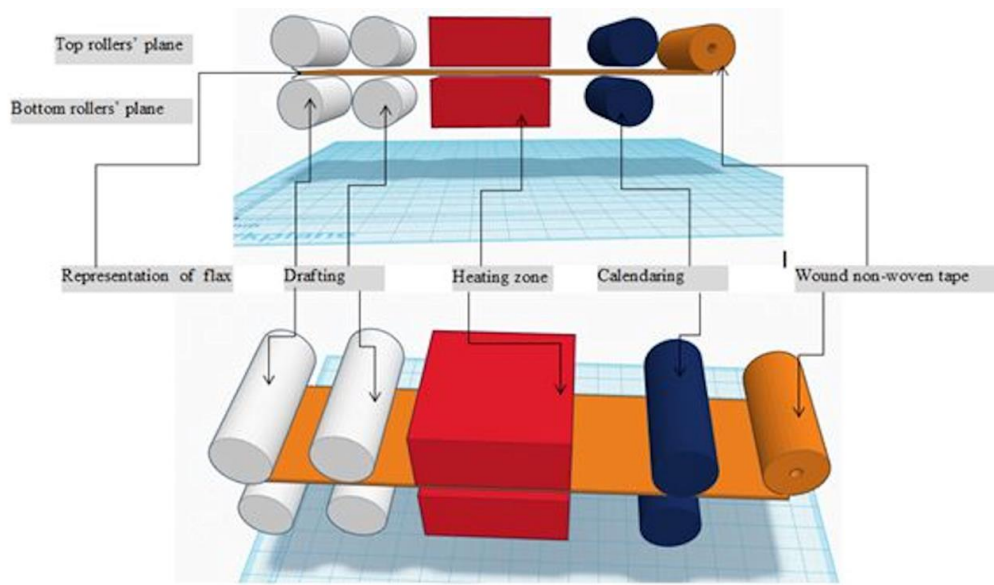

Figure 4    Schematic diagram of drawing, heating and condensing zones of a nonwoven tape manufacturing using Autodesk©.

## 2.3    Composite manufacturing and sample preparation

Cross-ply laminates were manufactured by the 'Vacuum Assisted Resin Injection Moulding' method using thermoset epoxy resin, LY564 as resin and Aradur 2954 (35% of the resin weight) as hardener from Huntsman. The curing was performed at 80 °C for 2 hours and the post-curing was completed at 120 °C for 6 hours, as shown in Figure 5.

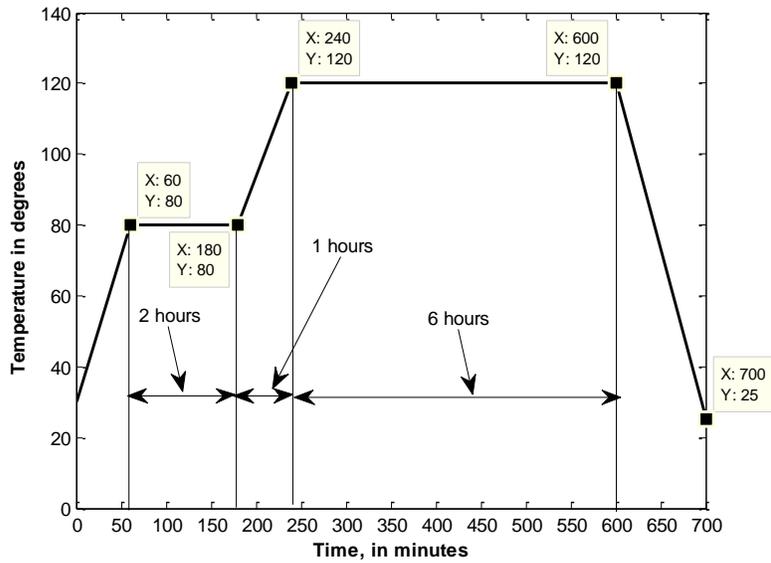

Figure 5   Curing cycle for LY564 and Aradur 2954.

Specifications of the tensile specimens are given in Table 2 . The specimens were prepared according to ASTM D3039. The fibre volume fraction (FVF) was calculated according to the ISO 14127:2008 [40], as described in Equation 1.

$$v_f = \frac{m_f/\rho_f}{V_c} \quad \ldots\ldots\ldots\ldots\ldots\ldots\ldots (1)$$

where,

$m_f$ = mass of the flax fibre,

$V_c$ = Volume of composite, and

$\rho_f$ = Density of fibre.

Table 2   Cross-ply laminate specifications

|  | Weave structure | Flax % | Gauge Length (mm) | No. of layers | Width (mm) | Thickness (mm) | Fibre volume fraction, % | Composite Density (g/cc) |
|---|---|---|---|---|---|---|---|---|
| Hopsack | Plain woven Hopsack | 88 | 150 | 3 | 247 | 2.64 | 39.64 | 1.26 |
| UD | Plain woven unidirectional | 84.53 | 150 | 6 | 252.9 | 2.89 | 36.89 | 1.25 |
| T180 | Nonwoven tape | 90 | 150 | 6 | 247.6 | 3.61 | 38.87 | 1.25 |
| GVT | Nonwoven tape with glass veil | 89.5 | 150 | 4 | 249.4 | 3.98 | 31.41 | 1.25 |

# 3   Results and discussion

Tensile tests were carried out according to ASTM D3039, where test coupons were rectangular, using an Instron 5982 machine with a 100 kN load cell. Gripping pressure was 50 bar and data captured rate was 200 milliseconds. The test data is compiled in Table 3.

Table 3   Mechanical properties of the cross-ply laminates

|  | Failure stress (MPa) | Failure strain (%) | Max load (kN) | 1st strain segment (%) | 1st modulus (GPa) | 2nd strain segment (%) | 2nd modulus (GPa) | FVF (%) |
|---|---|---|---|---|---|---|---|---|
| Hopsack | 72.97±1.32 | 1.32±0.07 | 4.76±0.1 | 0-0.5% | 7.91±0.26 | 0.5-1.31 | 4.21±0.24 | 39.64 |
| UD | 99.31±2.25 | 1.75±0.08 | 7.28±0.17 | 0-0.33% | 8.87±0.33 | 0.33-1.75 | 4.8±0.13 | 36.89 |
| T180 | 108.1±5.07 | 1.4±0.09 | 9.66±0.71 | 0-0.19% | 12.5±1.16 | 0.19-1.4 | 6.67±0.16 | 38.87 |
| GVT | 89.95±4.65 | 1.16±0.05 | 8.95±0.48 | 0-0.13% | 13.2±1.01 | 0.13-1.16 | 7.09±0.16 | 31.42 |

For comparative analysis of the composites of different fibre volume fractions, an extrapolation method using the rule-of-mixtures has been adopted here, which has been extensively used by previous researchers [28, 33, 36-38, 41]. The modulus of fibres inside the composite has been termed as 'fibre bundle stiffness', and the strength of fibres has been termed as 'fibre bundle strength' to expresses the stiffness and strength performances of flax fibres in the impregnated state. Equation 2 and Equation 3 below express the fibre bundle stiffness and fibre bundle strength respectively.

$$E_c = \eta_\theta \eta_l V_f E_f + (1 - V_f) E_m \quad \ldots \ldots \ldots \ldots \ldots \ldots \ldots \ldots \ldots \ldots (2)$$

where,

$E_c$ = modulus of composites, in GPa

$\eta_\theta$ = Krenchel fibre orientation factor which is

$$\eta_\theta = \sum_{n=0}^{1} a_n \cos^4 \theta_n$$

$a_n$ = Fraction of the fibres with the orientation angle $\theta_n$

$\eta_l$ = fibre length distribution factor

$V_f$ = Volume fraction of fibres

$E_f$ = modulus of fibres inside composites

$E_m$ = modulus of matrix

$\eta_l$ is considered here as 1 because the gauge length of all the specimens was 150 mm and the average length of flax technical fibres is 150 mm to 700 mm [42-44]. $\eta_\theta$ for the cross-ply structures (except Hopsack fabric) is 0.5 as exactly 50 percent of the total fibres in composites were laid up at an angle of 0° to the loading direction. For Hopsack, it was 0.4445 because the warp PPI (picks per inch) was 24 and weft PPI was 30. Therefore 44.45% of the total fibres were laid in the warp direction. Similar to the fibre bundle stiffness, fibre bundle strength has also been measured using the rule-of-mixtures that has been used by several researchers previously [28, 37, 45], as shown in Equation 2. The failure strain of the matrix, (4.5%) [46] is much higher than the failure strain of flax fibre (2%) [47]. Therefore, in this case, upon tensile loading, the failure of the flax fibres precedes matrix failure.

$$\sigma_c = (\eta \times V_f \times \sigma_f) + (V_m \times \sigma_m) \quad \ldots \ldots \ldots \ldots \ldots \ldots \ldots \ldots \ldots (3)$$

where,

$\sigma_c$ = strength of composites, MPa

$\eta$ = fibre orientation factor; for cross − ply = 0.5

$V_f$ = volume fraction of fibre

$\sigma_f$ = strength of fibre, MPa

$V_m$ = volume fraction of matrix

$\sigma_m$ = strength of matrix, MPa; (where,

$\sigma_m = E_m \times \varepsilon_c$; where in

$E_m$ = matrix modulus and $\varepsilon_c$ is composite failure strain)

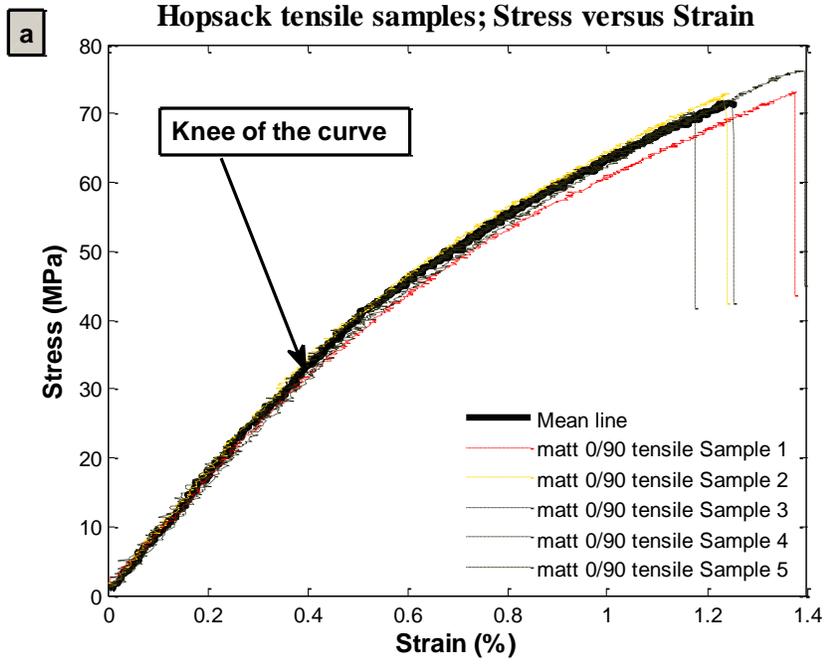
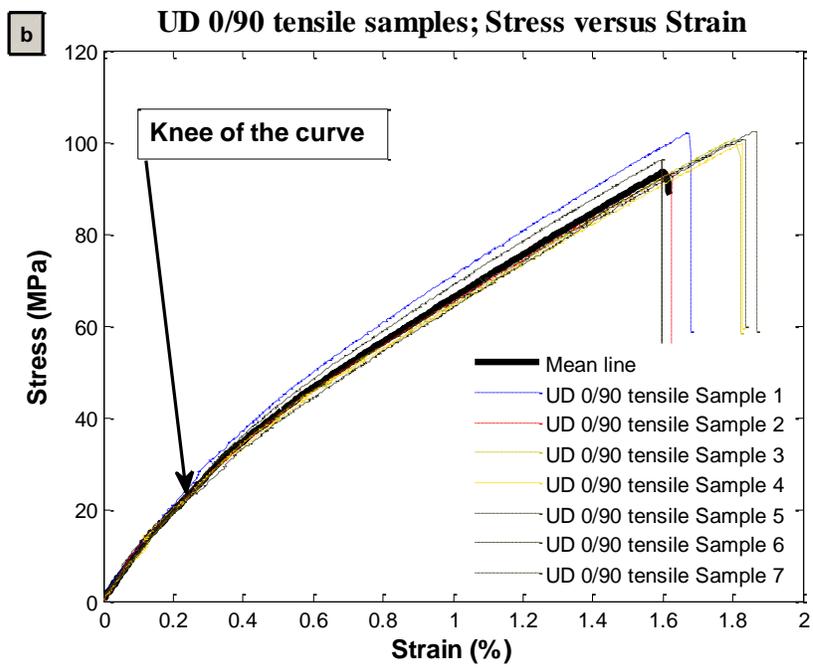

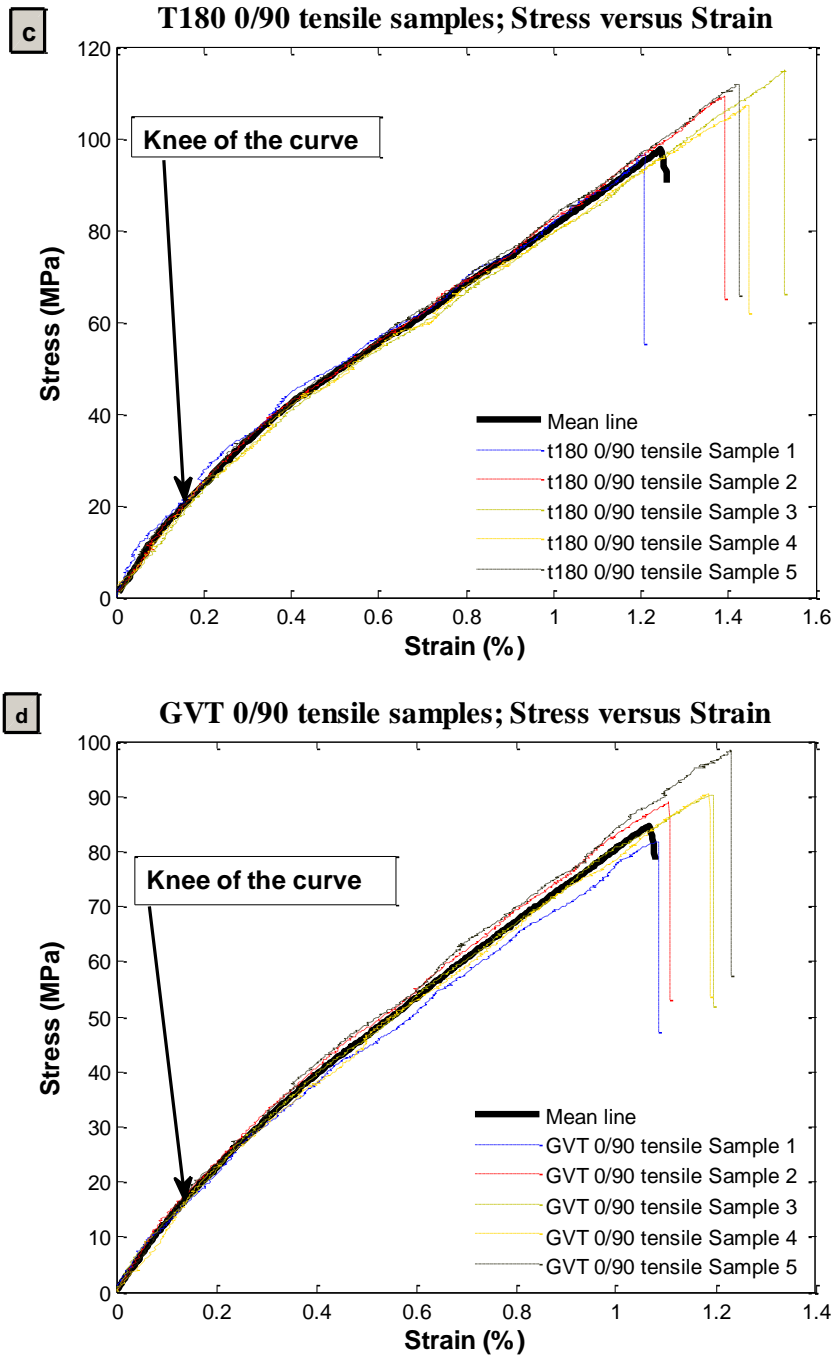

Figure 6  Mean stress-strain curve for: (a) Hopsack laminates; (b) UD laminates; (c) T180 laminates; and (d) GVT laminates.

Figure 6 shows the stress/strain curves for all the composites. On each plot, a distinct 'knee' point can be found, which indicates that each FFRC exhibited an initial higher modulus up to a certain strain limit, followed by a final degraded modulus. Previous researchers have also shown such stress-strain curves containing knee-points, wherein the tangent of the line below the knee-point was higher than the tangent of the line above the knee-point [33, 34, 41, 48].

The interpretation of the results, therefore, will focus primarily on the explanation of the onset of 'knee' in the stress-strain curves, followed by a comparative analysis of failure strain, failure strength, and initial modulus. Furthermore, the discussion will encapsulate the impact of fibre undulation on fibre morphology and how that undulation influences the corresponding composite's performances upon longitudinal tensile loading. The factors observed in this research that influence the composite's performances are: fibre undulation, individualisation of elementary fibres, homogeneity of fibre distribution, fibre pull-out, and impregnation of the lumens.

## 3.1 Onset of a 'knee' in the stress-strain curve

Though the occurrence of the disruption of the stress-strain curve of FFRC upon longitudinal tensile loading is reported in previous research works, no specific explanation about the reason for this phenomenon can be found in previous research works [33, 41]. Nevertheless, it is understandable that the incidence of this knee point is a factor dominated by the fibre morphology. In this research, some plausible reasons have been postulated based on the experimental observations. For the convenience of the discussion, the modulus of the line below the knee point of the stress-strain curve will be termed as 'modulus 1', and the modulus of the line above the knee-point of the stress-strain curve will be termed as 'modulus 2'. The values for moduli 2 have always been found to be lower than that for the moduli 1 for all the FFRC examined in this research.

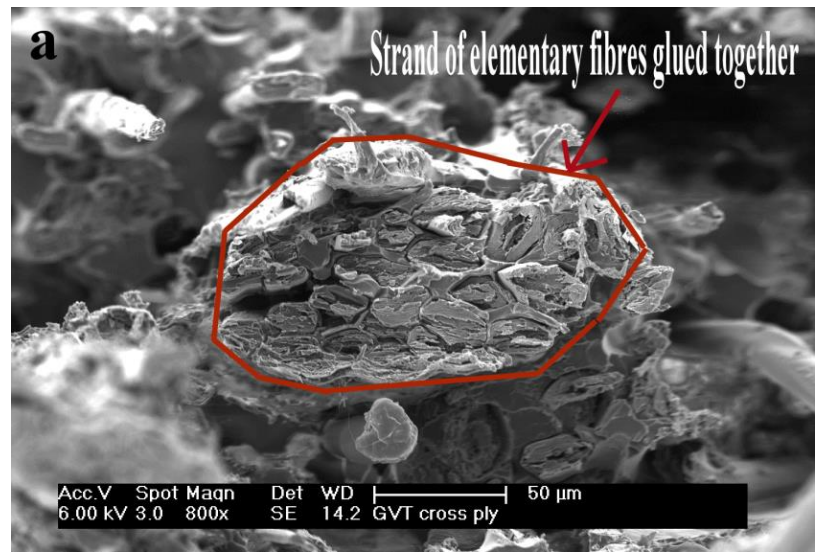

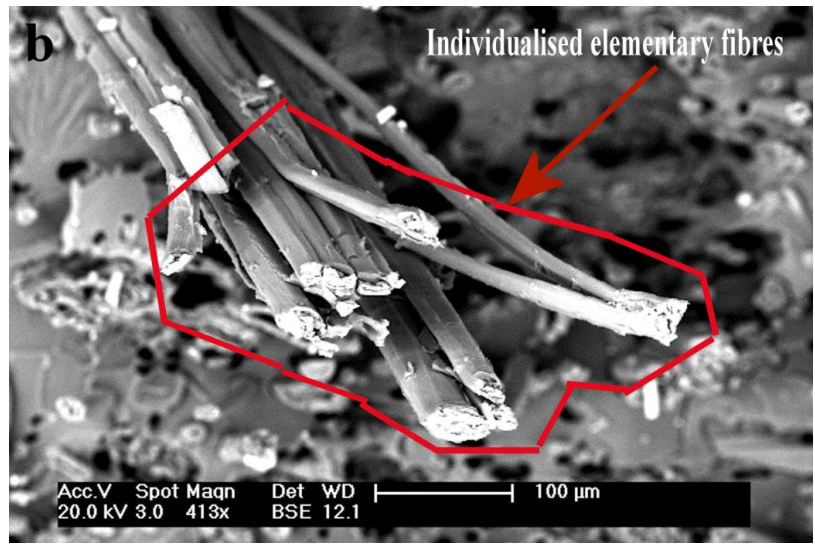

Figure 7  Fibres inside composite: (a) Elementary fibres are bonded with hemicellulose and pectin; and (b) Elementary fibres loose; without any natural matrix

It has been discussed earlier that a flax fibre functions as a 'technical fibre' – a bundle of elementary fibres, wherein, the elementary fibres are bonded together with naturally occurring hemicellulose and pectin matrices[21, 34, 37]. Technical fibres in fact resemble the structure and behaviour of a typical composite material. Hemicellulose and pectin work as matrix materials in this situation, and like other composite materials, the matrix of a technical fibre transfers stress upon tensile loading among the elementary fibres. When a FFRC is subjected to longitudinal tensile loading, the elementary fibres start to delaminate from pectin and hemicellulose, and at one point, the elementary fibres become completely separated and act like a dry fibre bundle (Figure 12). This is because, the reaction of a dry fibre in response to tensile loading is significantly different from that of an impregnated fibre. In a dry fibre bundle, inter-fibre friction and the strength of the weakest fibre dominate the failure[41]. But in a composite structure, the average strength of fibres is important as the matrix transfers stress from one stressed fibre to the adjacent fibres, or towards another part of a failed fibre. Stress transfer cannot occur for dry fibres, which implies that whenever the weakest fibre fails, a sudden drop in strength occurs, leading to a sudden failure. Figure 7 (a) shows elementary fibres bundled together, and Figure 7 (b) shows individualised elementary fibres without being cemented by hemicellulose and pectin. This failure of the matrix inside a technical fibre impedes the stress transfer amongst elementary fibres which probably reduces the modulus of a FFRC (modulus 2; above the knee-point) upon longitudinal tensile loading.

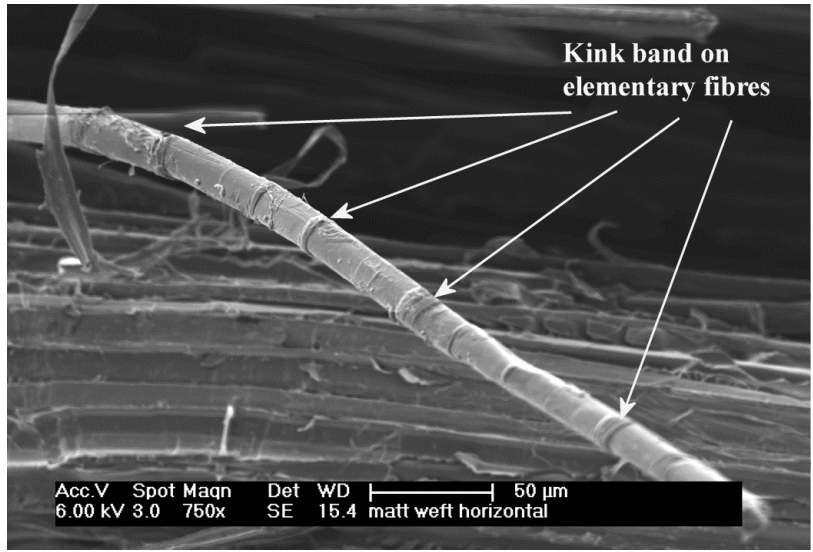

Figure 8  SEM image showing kink bands on elementary fibre surface.

Secondly, the kink bands of the flax fibres play a vital role in generating the reduced value of modulus 2 in comparison with the value of modulus 1. Figure 8 shows an elementary fibre with kink bands, taken from the damaged surface of a Hopsack laminate after tensile loading. The places, where kink bands exist act as stress accumulators. This stress concentration around the kink bands initiates crack propagation, which can lead to the commencement of modulus 2 (the reduction of the gradient in the line above the knee of the stress-strain curve of FFRC) upon longitudinal tensile loading. Furthermore, at the same time, kink bands are also susceptible for fibre failure within the technical fibre bundles upon tensile loading [27, 31].

Occurrence of a knee point during tensile loading for dry flax fibre bundles has also been observed by several researchers [25, 31]. A notable point is that, for a dry flax fibre bundle subjected to tensile loading, strain hardening (straightening out of the kinked regions by reorientation of the micro-fibrils towards the fibre axis) occurs; this in turn increases the gradient of the line above the knee-point, unlike the stress-strain curves of FFRC. Inside a composite, the kink bands are kept locked by the matrix, which is not the case for the dry fibres without impregnation. Therefore, the occurrence of strain hardening for flax fibre inside a composite is much less likely, as the process is hindered by the matrices.

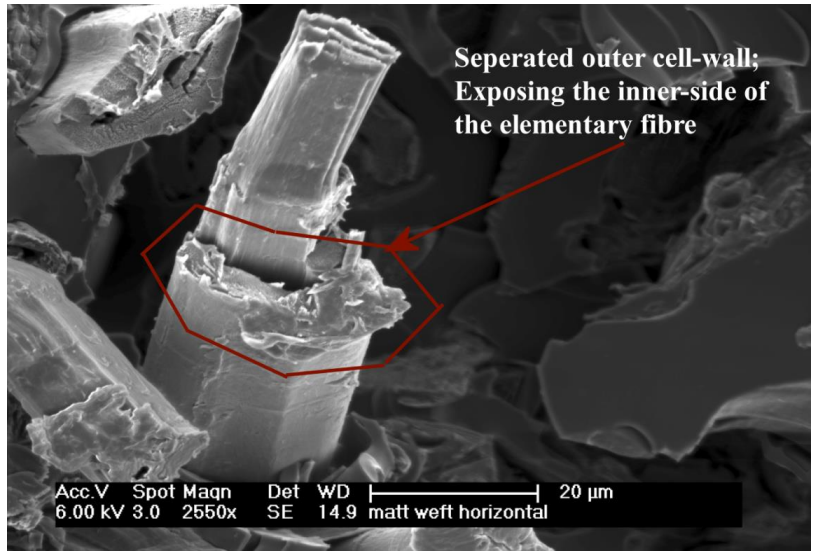

Figure 9  Detachment of the outer cell wall from the inner cell wall of a flax fibre.

Thirdly, the outer cell wall of an elementary flax fibre acts as a 'segmented sleeve' that shows a certain degree of relative lengthwise movement[25]. During a gradual increment of load, the cell walls may slip over each other, which in turn may cause complete failure of the outer cell wall, leaving the inner part empty. At that point, the elementary fibre with the completely severed outer cell wall cannot bear as much load as it could with an intact structure. Kersani et al. (2014) have mooted this reason in their work for the modulus 2 of the stress-strain curve of FFRC upon longitudinal tensile loading, but they presented no supporting evidence. In this research, the failure of the outer cell wall has been supported by SEM micrographs, as shown in Figure 9. It may be seen from Figure 9 that the outer cell wall has been detached from the inner cell wall, which leaves the inner cell wall exposed, without any contact with the matrix of the composite. Thus, upon longitudinal tensile loading, the failure of the outer cell wall can result in a lower tangent value of the line above the knee point of the stress-strain curve of a FFRC.

## 3.2  Strain of different structures at which the knee point commences

From Table 3, it may be observed that the woven structures, namely the Hopsack (modulus 1 region: 0 to 0.5% strain) and the UD fabrics (modulus 1 region: 0 to 0.33% strain) demonstrate a higher range of strain percentages for the modulus 1 region than that of the nonwoven structures (T180: 0 to 0.19% strain for modulus 1, and GVT: 0 to 0.13% strain for modulus 1). The Hopsack fabric consists of undulating warp yarn along its length. Therefore, upon tensile loading up to a certain strain percentage, the structure straightens up its crimped warp yarns. At the same time, delamination between elementary fibres and natural occurring matrices (pectin and hemicellulose that bundle the elementary fibres into technical fibres) begins. The constituent undulations of the Hopsack structure result in a higher strain percentage for the onset of the knee point than that in the UD, T180, and GVT. It should be noted that Hopsack warp yarn exhibits 17.7° crimp in dry fabric, whereas the unidirectional yarns of the UD fabric exhibit only 6.1 ° crimp in dry fabric.

The reason for the nonwoven composites exhibiting a lower strain percentage at the onset of the knee point, in comparison with the woven structures, is the fibre undulation. Neither in the dry state nor in the composites, have the fibres of the nonwoven structure showed any discernible undulation. Therefore, upon tensile loading of the composites, the laminates do not show any extra extension before the fibres start to experience stretching.

T180 laminates show slightly higher tensile strain values in comparison with GVT laminates. This happens because the GVT fabric contains a glass interleaf (with randomly oriented short glass fibres).In a four-layer GVT laminate, the glass-interleaf adds up to a significant proportion of the total laminate volume; 11.5% of the total weight of a sample, on average. This randomly orientated portion causes an early failure of the GVT structure and also an early onset of the knee-point (hence, an initiation of modulus2) in the stress-strain curve upon tensile loading, compared to that in the T180 structures.

### 3.3 Analysis of the failure strain

Table 3 shows the failure strain values of all the cross-ply laminates. Between the two woven fabric laminates, UD shows a 32% higher extension than Hopsack. This occurs because of the intrinsic crimp of the Hopsack structures. A reinforcement yarn without any out-of-plane undulation can bear tensile loading more efficiently than the yarns with undulation [49, 50]. During tensile loading, the undulated warp yarns of the Hopsack fabric first tend to stretch themselves out towards the loading direction. As the extent of the crimp is much higher in the Hopsack structures, before the warp yarns become fully straightened, the matrix cracking is initiated. Thus, it is the yarn waviness that reduces the load bearing efficiency of the Hopsack laminates. It can be seen from Table 3 that the UD laminates show failure strain (failure strain averaging 1.74%) as 32% higher than that of the Hopsack laminates (failure strain averaging 1.32%). In addition, it has been discussed earlier that the manufacture of Hopsack fabric involves more undulation of the constituent yarns than for component yarns of the UD fabric, hence generating more kink bands in the constituent fibres of the Hopsack fabric. As kink bands work as stress accumulators [26, 31], failure takes place earlier in the Hopsack structures than in the UD structures. Between the two nonwoven structures, T180 (failure strain averaging 1.4%) has shown to resist 21% more strain than that of the GVT (failure strain averaging 1.16%) before failure, because of the absence of glass veils (containing randomly oriented glass fibres) in the former structure.

Fibre pull-out is another phenomenon that can be attributed to the differing failure strains of different structures. Fibre pull-outs occur when a discontinuous fibre is embedded in a relatively tougher matrix. Flax fibres contain kink bands along their length which implies that whilst embedded in matrix, a flax elementary fibre in effect exists as 'segments' of 'short fibres', 'joined together' [31]. Thus, a kink band acts as a weak link in an elementary fibre and effectively mitigates the continuity of an elementary flax fibre. These weak links may not exist in the same plane as a composite fracture. When a fibre breaks, it introduces stress concentration into the matrix. This stress concentration may be relieved as matrix yielding also takes place. Therefore, along the embedded length of an elementary fibre, matrix cracking may not occur even though there are fibre breakages at the kink bands. In such a scenario, the broken fibre may be pulled out of the matrix rather than failing again at the plane of the composite fracture [49].

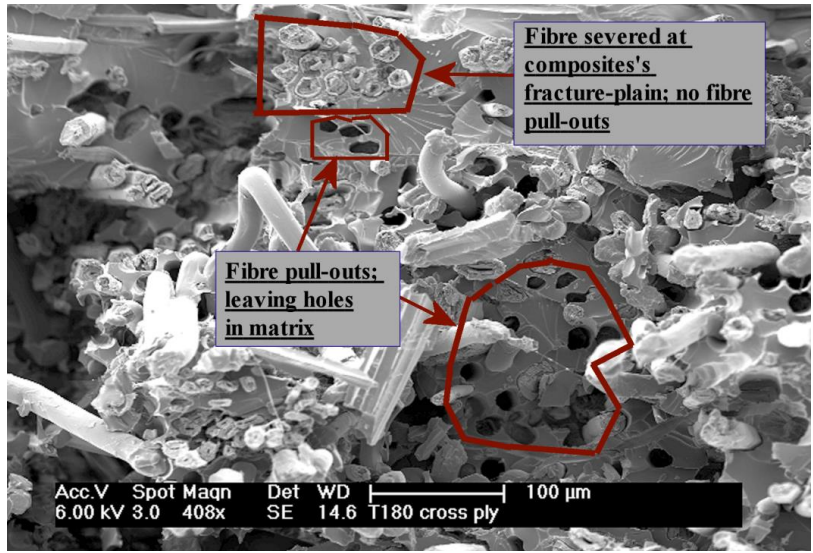

Figure 10   Fibre pull-outs; T180 specimen damaged from tensile test.

Figure 10 shows a typical example of an occurrence of fibre pull-out, wherein the pulled-out fibres have left the holes in the matrix at the plane of the composite failure. The specimen shown in the photomicrograph is a T180 specimen damaged during tensile failure. Figure 10 also illustrates an example, where fibre pull-out from matrix and fibre-breakage at the composite's fracture-plane have occurred simultaneously.

If in a composite structure, fibre pull-outs occur at a greater extent, then the failure strain percentage of the composite will be higher [34]. From Table 3, it can be found that the failure strain of the UD laminate (1.75%) is 25% higher than that of the T180 structure (1.4%). Neither the UD nor the T180 show any apparent waviness in their composite structures. Therefore, the difference in the failure strain percentage of these two structures can be directly linked with the fibre pull-out phenomenon. Comparing the T180 and the UD specimens, fibre individualisation has been found to be higher for the former. However, as the UD composites are composed of yarns, rather than the individualised elementary fibres (as found in the T180 structures), the strand of fibres in the former structure act as monolithic units. Generally, for two composite structures with different proportions of individualised fibres, fibre pull-out will occur to a greater extent in the structure with a higher proportion of individualised fibres. This is evident from a comparison of Figure 13 (a) and Figure 13 (b), wherein, the occurrences of fibre pull-out in the former are higher than in the latter. However, in this case, the structural differences between the UD and the T180 specimens are important. As the UD material is manufactured from yarn, the length of a yarn pulled-out from within the matrix surface (in UD structures) is greater than the average length of the elementary fibres in the T180 structures. Therefore, though the occurrence of fibre pull-out is higher in T180 (due to the greater number of individual elementary fibres in T180 specimens, in comparison with UD), the effect of fibre pull-out (which influences the failure strain) is higher in the UD structures. This usefully explains why the failure strain for UD composites is 25% higher than that of T180 structures.

## 3.4 Analysis of the composite strength

Table 4 compiles the tensile strength results at a 95% confidence level, recorded from tensile tests executed on an Instron 5982 machine, and shows the extrapolated data created using Equation 3. It can be observed that the average fibre bundle strength of Hopsack (296.56±8.38 MPa) structures is 29% lower than that of the UD (382.73±6.02 MPa) structures. The main cause that can be attributed to this difference is the differing extent of the undulation of the fibres within the structures. Both the fabrics have been manufactured using same yarn. However, the yarns used in the Hopsack laminates have shown a 14° undulation inside the composites whereas the UD yarns did not demonstrate any waviness. It has been discussed earlier that the undulation increases the number of kink bands in the flax fibre, which in turn lowers the composite's strength. In addition, as a principle of mechanics, an undulating reinforcement is unable to bear load as much as a straight reinforcement can.

Table 4    Strength of the laminates

|  | Failure stress (Mpa) | Failure strain (%) | FVF (%) | Fibre bundle strength (Mpa) | Normalised FVF (%) | Failure stress at 40% FVF (Mpa) |
|---|---|---|---|---|---|---|
| Hopsack | 72.97 | 1.32 | 39.64 | 296.56±8.34 | 40 | 73.32±7.29 |
| UD | 99.31 | 1.75 | 36.89 | 382.73±6.02 | 40 | 103.85±2.25 |
| T180 | 108.1 | 1.4 | 38.87 | 441.72±15.23 | 40 | 110.18±5.06 |
| GVT | 89.95 | 1.16 | 31.42 | 440.91±14.61 | 40 | 106.28±4.65 |

The main difference among the UD, T180, and GVT is the geometry of their reinforcement assemblies. UD is warp-knitted fabric made from twistless flax yarns, whereas T180 and GVT are nonwoven tapes, made from technical fibres. During manufacture, the fibres of tapes undergo additional drafting, which minimises the kink bands to some extent by strain hardening [25, 31, 34, 51]. Figure 11 shows the images of fibres under polarising filter. It may be observed that the number of kink bands along a certain fibre length is higher in a fibre taken from the T180 tape, shown by Figure 11 (a) than in a fibre taken from the UD fabric, shown in Figure 11 (b).

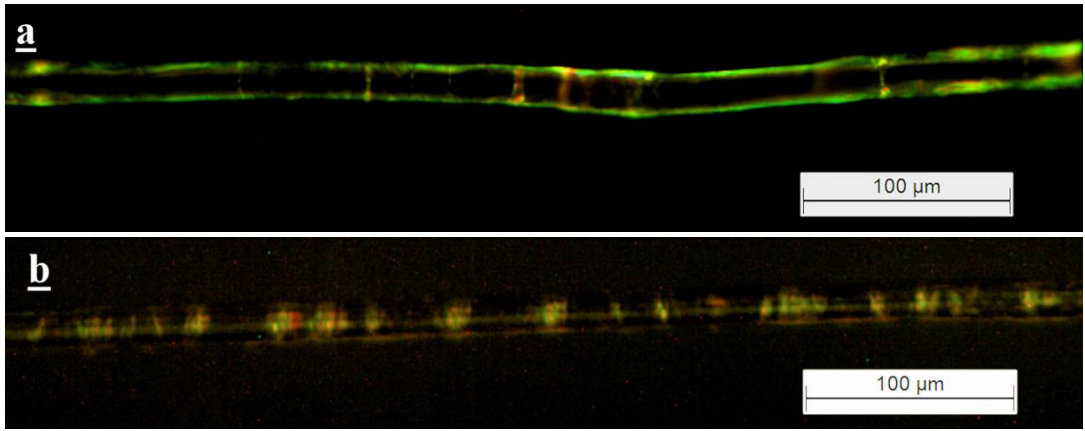

Figure 11  Optical microscopic image with polarisation filer; Kink bands on: (a) T180 fibre; and (b) UD fibre.

Secondly, individualisation of the elementary fibres plays a key role in achieving higher strengths for the tape-reinforced composites in comparison with the UD laminates. Fibres in the UD fabric remain as strands of fibres clustered into yarns, whereas tapes are assemblies of loose fibres. During manufacturing, nonwoven tapes undergo an additional step of drawing and drafting to ensure better individualisation and parallelisation of the constituent fibres. During that drafting, a significant proportion of the pectin and hemicellulose bonding of the elementary fibres inside the technical fibres may suffer damage, and a delamination between elementary fibres and the naturally occurring hemicellulose and pectin matrices may take place. A study was conducted as part of this research wherein the average fibre length of the fibres from raw sliver and twistless yarn (drafted sliver) was captured. It was found that the median length of the fibres from the raw sliver was 119.37±4.69 mm, and from the drafted sliver, it was 98.94±4.02 mm. These values indicate that the drafting operation imparts axial force onto the fibre strand which in turn individualises elementary fibres from the technical fibre bundle. In contrary, the fibres of the UD structures remain clustered as yarn inside composites.

Due to this additional drafting during tape manufacture and fibre arrangement of the tapes, the presence of individualised elementary fibres in the tape-reinforced composites is higher than that of the UD composites. This increased individualisation of the fibres in the tapes renders a better homogeneity of fibre distribution and a reduced incidence of resin rich areas. Bos et al. (2002)[52] mentioned in their research that the technical fibre strength is 57% of the corresponding elementary fibre strength. So it can be inferred that if the fibres in the laminate are arranged in such a way that the technical flax fibres are broken into elementary fibres, the strength will be enhanced. If the elementary fibres remain as individual units instead of being bundled during composite manufacture (i.e. technical fibre), they can be fully encased by the matrix and can demonstrate superior mechanical properties. This happens primarily due to the bundle effect [53], and secondly, individual elementary fibres offer more surface area to the encapsulating matrix, which a technical fibre (i.e. a bundle of elementary fibres) cannot do up to the same extent if those were individualised. A greater surface area ensures better stress transfer between fibre and matrix. Inside a composite, the role of the resin is to transfer the stress from the reinforcement. This is why better interfacial bonding ensures better tensile properties and better stress transfer [49,50]. If dry fibres remain as technical fibre bundles inside a composite, then the internal elementary fibres within a technical fibre during tensile loading may separate and may remain as dry strands of fibres within the composite structure.

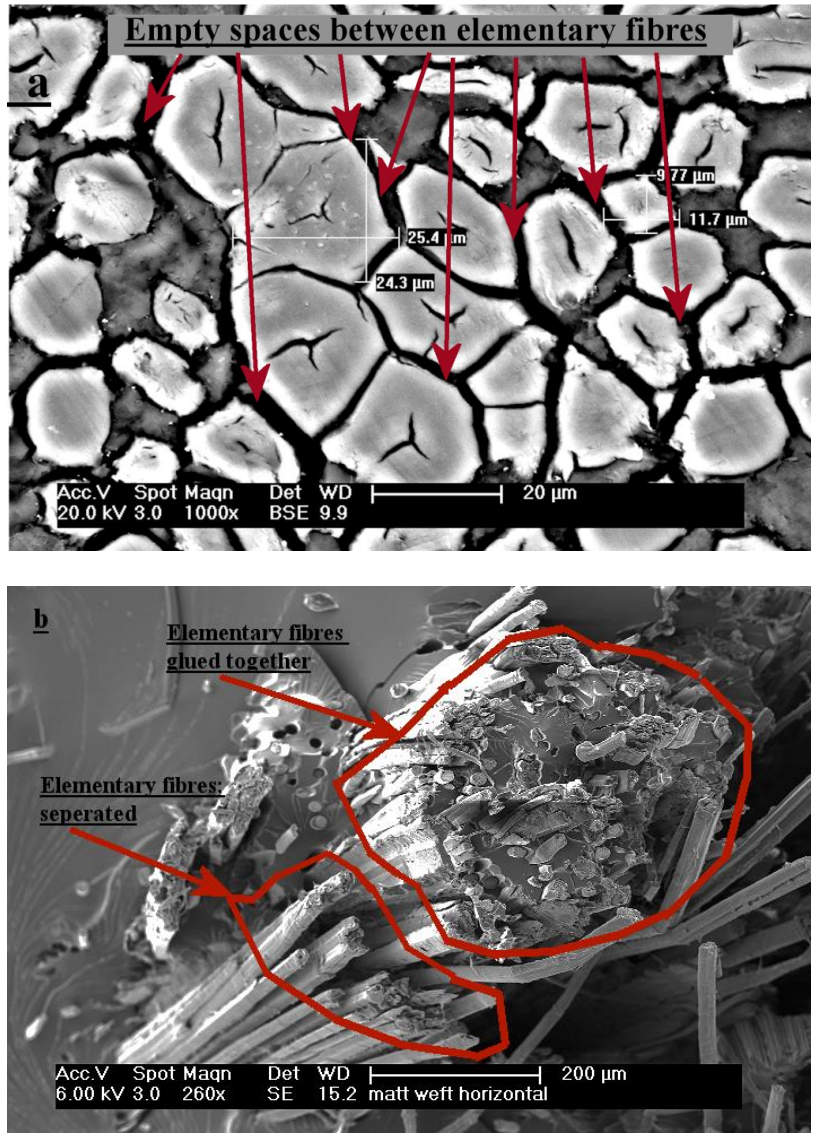

Figure 12  Gaps between elementary fibres in: (a) UD laminate; and (b) Hopsack laminate.

Figure 12 (a) shows a bundle of elementary fibre having empty spaces in between the fibres. Therefore, the separated elementary fibre will not be able to transfer its strength to the matrix. Figure 12 (b) shows empty spaces between two elementary fibres, which leave an elementary fibre partially dry along its length within a composite. On the contrary, Figure 14 (a) shows elementary fibres being surrounded with resin for a GVT laminate. Both Figure 12 (a) and Figure 14 (a) represent cross-sectional images of undamaged flat surfaces of a UD and a GVT composite respectively. If a significant number of technical fibres exist in a composites structure instead of that of individual elementary fibres, there will be significant numbers of fibres without any contact with resin, which will in turn lower the strength. As the presence of individual elementary fibres is higher in the tapes than in the UD fabric, the strength of UD composites is lower than that of the tape-reinforced composites.

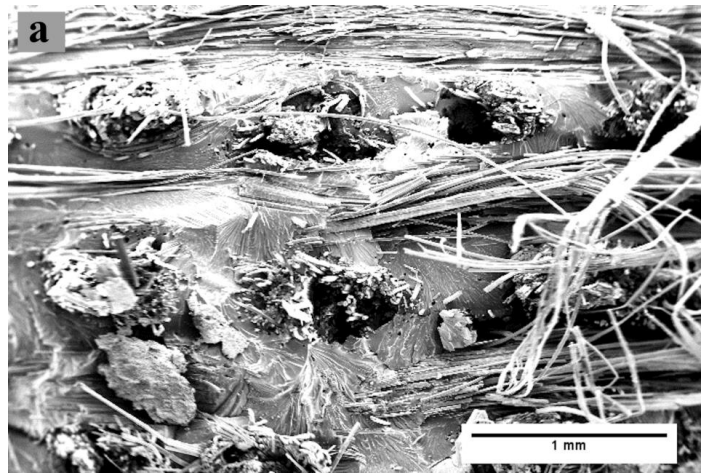

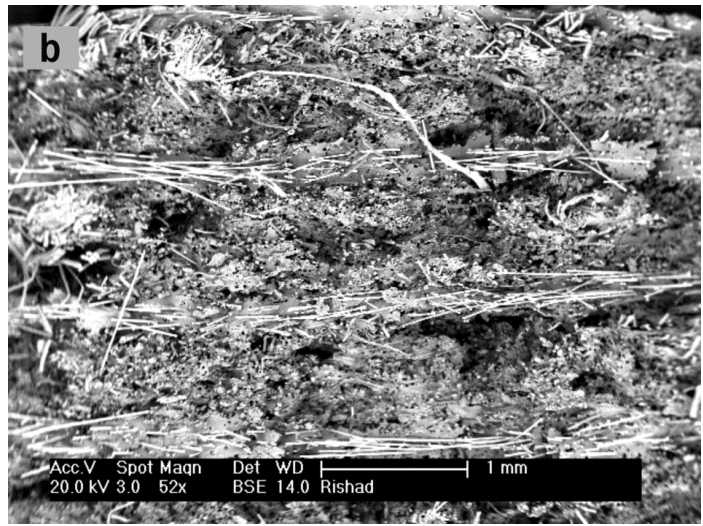

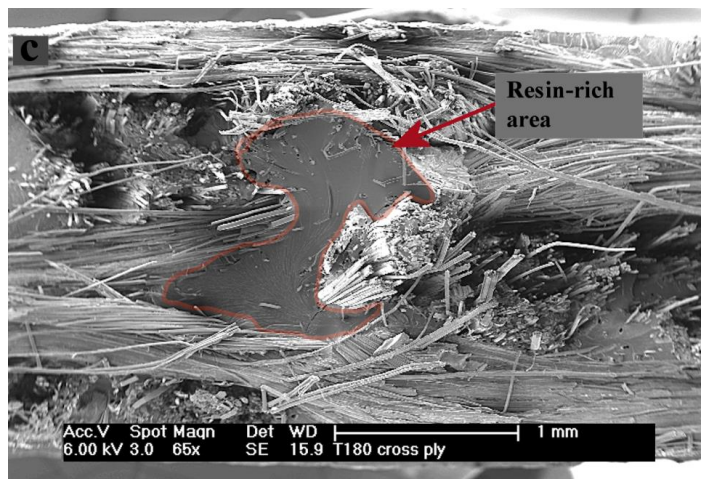

Figure 13  Top-view of composite sample, ruptured after tensile test: (a) UD; (b) T180; and (c) Hopsack.

Thirdly, a greater extent of individualisation of elementary fibres ensures a more homogeneous distribution of the fibres inside the composites. Figure 13 shows a comparative scenario of the fibre homogeneity in flax composites and also the extent of resin rich areas. A homogenous distribution of the fibres across a composite structure results in fewer resin rich areas, which is beneficial for the mechanical performance of the composite.

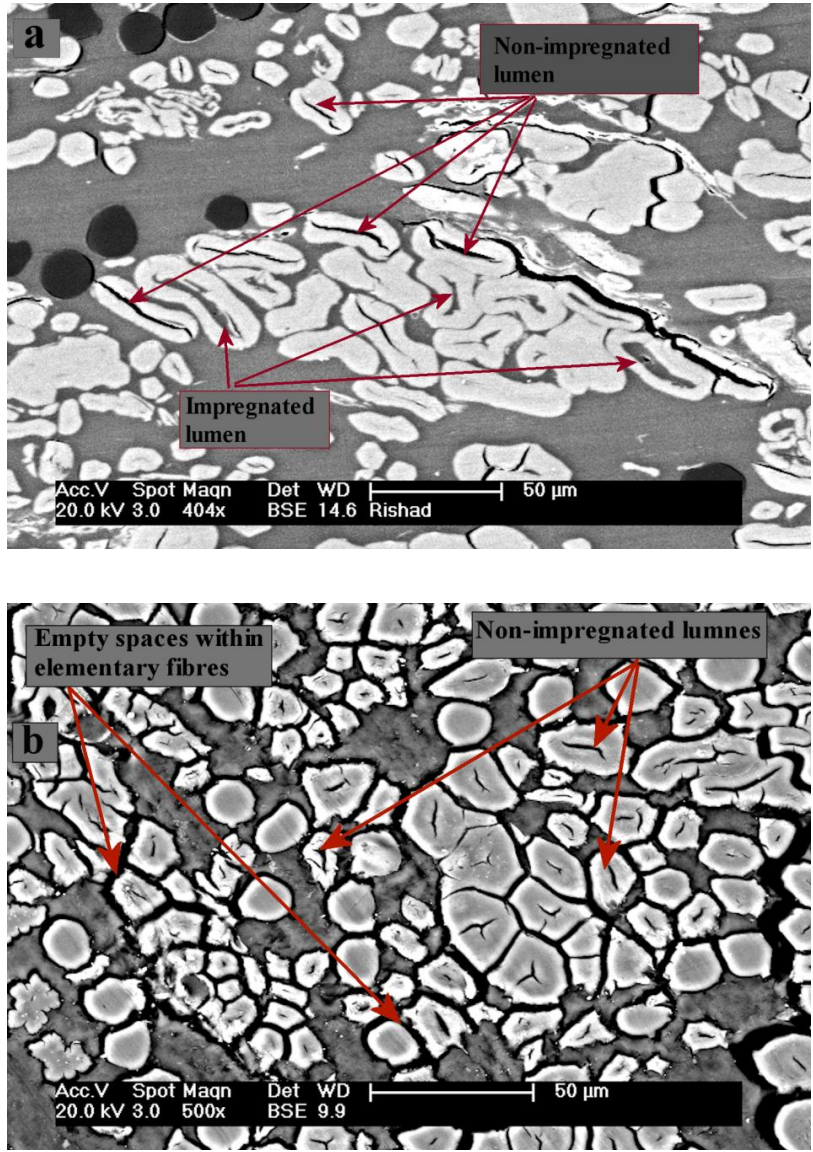

Figure 14  Flat surfaces of undamaged composite specimens: (a) GVT; and (b) UD.

Finally, another phenomenon can be attributed to the superior strength properties of the T180 and GVT composites in comparison with the UD, and this is the impregnation of the lumens of the elementary fibres. Figure 14 (a) and Figure 14 (b) show examples of flat surfaces of undamaged composites. From both the figures, it can be seen that the presence of the impregnated lumens are greater in the GVT laminates in comparison with the UD laminates, consequently, the latter exhibits lower fibre bundle strength. If the lumens are infused too, that will add additional stiffness to the composites. This will happen because more of the fibre surface area will be exposed to the resin, which will improve the stress transfer between matrix and fibre. Moreover, during tensile loading, at the point when the pectin/hemicellulose matrices will start to suffer damage inside a technical fibre, the elementary fibres will be separated from each other and will remain as individual dry fibres, resulting in rendering negative influence for the strength of composites. This phenomenon has been demonstrated in Figure 12 (b) by a Hopsack laminate damaged during tensile testing. The impregnated and non-impregnated lumens can also be observed from Figure 15 (a) and Figure 15 (b).

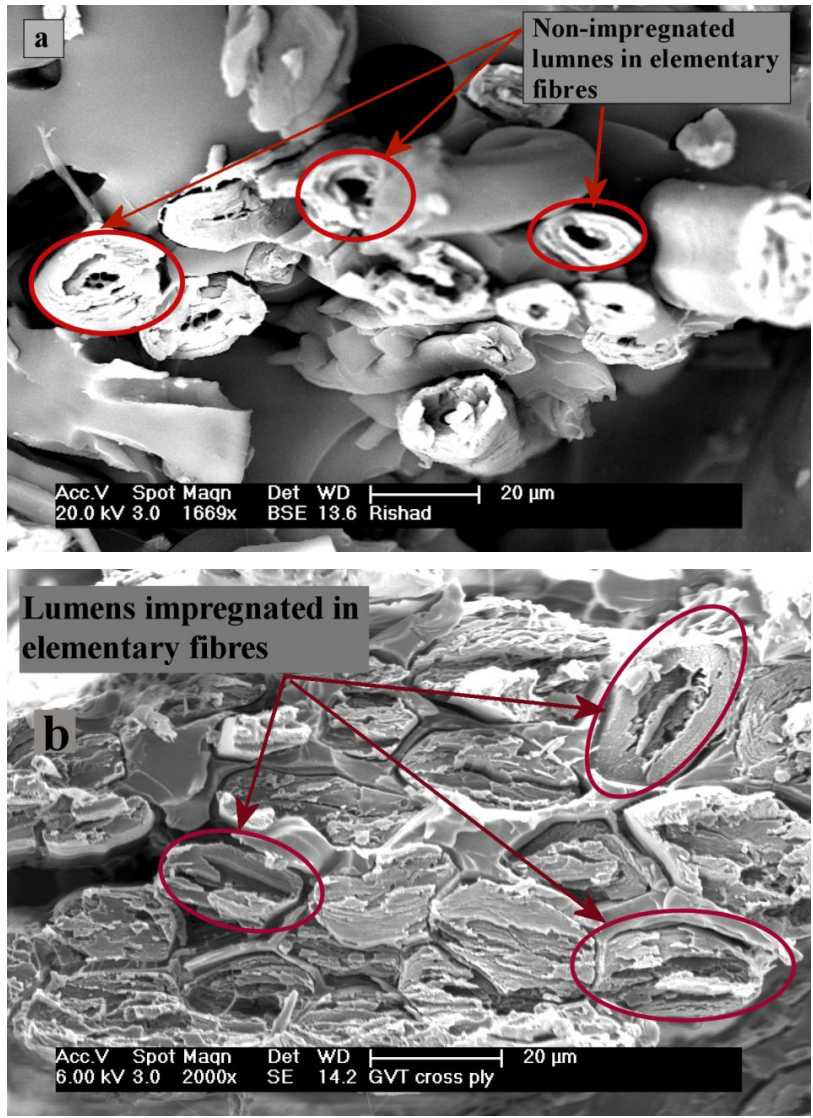

Figure 15   Damaged specimens with empty and impregnated lumens; (a) UD; and (b) GVT.

It may be observed that the fibre bundle strength values for GVT and T180 are almost equal. In GVT, an extra 2D glass veil has been inserted and this imparts a hybridisation effect on the composites. Attaching the glass veil increases the modulus; however, it does not confer any significant improvement in respect of strength properties.

## 3.5 Analysis of modulus

Table 5 Calculated fibre bundle stiffness of the cross-ply laminates

|  | 1st strain segment (%) | 1st modulus (GPa) | Fibre bundle stiffness (GPa) | 2nd strain segment (%) | 2nd modulus (GPa) | Fibre bundle stiffness (GPa) | Fibre Volume Fraction (%) |
|---|---|---|---|---|---|---|---|
| Hopsack | 0-0.5% | 7.91±0.26 | 35.98±1.18 | 0.5-1.31 | 4.21±0.24 | 14.99±0.85 | 39.64 |
| UD | 0-0.33% | 8.87±0.33 | 39.19±1.46 | 0.33-1.75 | 4.8±0.13 | 19.27±0.52 | 36.89 |
| T180 | 0-0.19% | 12.5±1.16 | 56.14±5.21 | 0.19-1.4 | 6.67±0.16 | 26.14±0.63 | 38.87 |
| GVT | 0-0.13% | 13.2±1.01 | 72.67±5.56 | 0.13-1.16 | 7.09±0.16 | 33.78±0.76 | 31.42 |

Figure 16 demonstrates the average stress/strain response for up to 0.2% strain, for all four types of laminates used in this research, wherein the nonwoven tape reinforced composites show greater modulus value in comparison with the composites reinforced with UD or Hopsack fabric. As the fibre volume fractions of the laminates were different, Equation 2 has been applied to derive a normalised comparison. Table 5 contains a compilation of the full set of test results at the 95% confidence level, captured using an Instron 5982 machine.

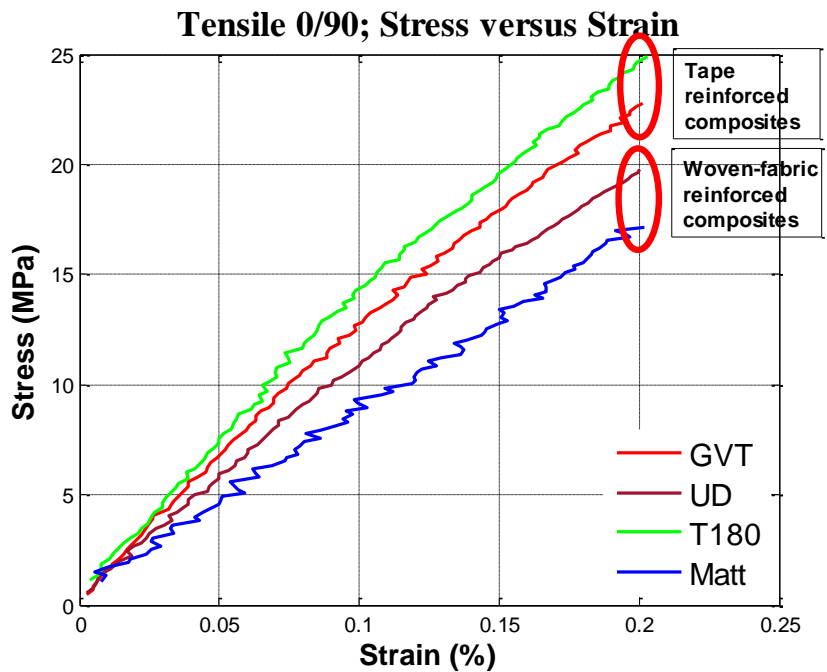

Figure 16 Stress-strain curve at initial stain%.

The stiffness of flax fibre reported by various researchers lies in the range between 12 and 85 GPa, and in terms of strength, it is between 600 and 2000MPa [25, 44, 54-57]. GVT shows here the highest result wherein the initial fibre bundle stiffness calculated by Equation 1 is 72.67GPa, which is the highest among the currently published research papers that discuss unidirectional flax fibre reinforced epoxy composites [28, 33, 36, 38, 41]. T180 shows 56.14 GPa which is also higher than the results reported by Bensadoun et al. (2017). Hopsack returns the lowest value of 35.98 GPa which is almost 51% of the highest performance shown by GVT. T180 and UD are respectively the second and third highest. The method of calculating fibre bundle stiffness used in this research has also been followed by previous researchers [28, 33, 41]. The main focus during the production of nonwoven tape in this research was to maintain as good fibre alignment as possible. Baets et al. (2014) conducted similar research in which they used aligned flax fibres infused with epoxy resin[51]. In their research, stiffness of hackled flax reinforced composites were returned as of 62.9 GPa (using Equation 2), composites constructed using roving reached 51.4 GPa[51], and those constructed from yarn attained 43.1 GPa. In this current research, the results for GVT were found to be 72.67GPa; for T180, the value was 56.14 GPa, and for UD laminate the stiffness is 39.19 GPa. All the values appear to be higher than the results reported by Baets et al. (2014). Moreover, in this research, cross-ply laminates have been used whereas previous literature describes unidirectional laminates. Therefore, it can be claimed that composites reinforced with nonwoven tapes produced in this research offer superior mechanical properties to existing similar products on the market.

The variation of the moduli of different structures reflects changes in one major variable in the construction of the cross-ply laminates: fibre undulation. Out-of-plane undulation of the fabric reinforced composites has a negative impact on their longitudinal tensile and compressive properties [13, 15, 16, 19]. Hopsack and UD fabrics contain 17º and 6º fibre undulation whereas nonwoven tapes contain no discernible crimp. In comparison with Hopsack and UD, UD shows 9% higher modulus values which can be attributed to the undulation of the fibres of the Hopsack. In composite, Hopsack shows 14 º undulations whereas UD exhibits no apparent crimp.

The fibre bundle stiffness of T180 for modulus 1 region was found to be 56.14 GPa, which indicates a very good translation of fibre stiffness properties into composite stiffness. Neither UD nor T180 show any measurable crimp in their composite structures. However, T180 shows fibre bundle stiffness values 43% higher than that of the UD in the modulus 1 region. The reasons underlying the superior performance of T180 laminates are the reduced incidence of kink bands, improved homogeneity of fibre distribution, more successful individualisation of elementary fibres, and better impregnation of the lumens of the elementary fibres. It is notable that the GVT structures show the highest fibre bundle stiffness values in the modulus 1 region amongst all the structures, 29% higher than that of the T180 laminates. The interleaved glass webs inserted into the GVT fabric structure has provided the additional initial modulus measured in the GVT laminates.

The Young's modulus values for the modulus 2 regions, as shown in Table 5, follow the same order as is in the modulus 1 region. For the modulus 2 region also, GVT composites exhibit the highest stiffness values and Hopsack composites demonstrate the lowest stiffness values. UD and T180 are third and second from the top respectively. Amongst recent publications, only Bensadoun et al. (2017) have shown that for unidirectional laminates, the modulus in modulus 2 region is 46.1 GPa, which is higher than the highest modulus found for the modulus 2 region (for GVT: 33.78 GPa) in this research. However, the laminates used in this work are cross-ply, which would be expected to show inferior tensile properties than unidirectionally laid specimens.

It has been discussed earlier that due to the intrinsic morphological structure of the flax fibre, the Young's modulus drops after the application of a certain strain percentage. The decrease of modulus in percentages between modulus 1 and modulus 2 regions for each composite structure have been found to be almost similar, and listed as: Hopsack 58.33%, UD 50.8%, T180 53.43%, and GVT 53.52%. It can, therefore, be postulated that the degree of the drop of modulus is not a function of the structural crimp, rather it is a phenomenon related to the fibre's morphology.

# 4  Conclusion

A novel approach has been deployed in this research to manufacture flax nonwoven tape with minimal fibre waviness, in order to increase the longitudinal tensile performances of FFRC. Crimp also increases the incidence of kink bands to the elementary flax fibres, which reduces the translation efficiency of the fibre to the composite. To eliminate the fibre undulation, a nonwoven tape namely T180 (blended with 10% PLA) and a glass veiled tape (GVT; formed with a glass interleaf) were manufactured, wherein the fibres remained in a highly aligned position. These tape-reinforced composites (with no discernible waviness) were compared, in terms of longitudinal tensile properties, with plain woven Hopsack fabric (14º fibre waviness, and wave amplitude to wave length ratio 0.14) and warp-knitted UD fabric reinforced composites.

Each stress-strain curve upon tensile loading showed a distinct knee point, which is an intrinsic feature of FFRC. Three reasons have been mooted in this research that might have influenced the onset of the 'knee- point' and the decrease in the modulus of the line above the knee-point. First reason was the delamination of the elementary fibres from the naturally occurring hemicellulose/pectin matrices upon tensile loading. Hemicellulose and pectin act as matrices to bond elementary fibres in technical fibres. Second reason was the generation of cracks due to the presence of kink bands that initiated damages in the epoxy matrix. Thirdly, the failure of the outer layers of the elementary fibres during tensile loading that caused a decrease in resisting the tensile load. Kersani et al. (2014)[48] also speculated that the failures of the outer layers of the elementary fibres under tensile loading was liable for the occurrence of the 'knee-points' in the stress-strain curve of their composites. However, in this research, this phenomenon was supported by SEM micrographs and analysed qualitatively.

For the failure strains, UD has shown a 32% higher value in comparison with that of Hopsack. The reduced failure strain value of hopsack can be attributed to the higher crimp percentage of its constituent yarns. Due to this higher crimp in the constituent fibres in Hopsack structures, matrix yielding was initiated in the Hopsack structures at a lower strain percentage than that in the UD structures, resulting in lower failure strain value for the former. Between the two nonwoven structures namely GVT and T180, presence of the randomly oriented glass veils was liable for the GVT to exhibit a lower failure strain value than that of the T180 structures. Between T180and UD structures, the effect of fibre pull-out was liable for the UD to have a higher failure strain value. T180 contained more individualised elementary fibres, hence experienced higher fibre pull-outs, whereas UD is comprised of yarns with bundled fibres. Thus a yarn pulled-out from within the matrix of a UD structure was greater in length than the average length of the elementary fibres that were pulled out in a T180 structure. Pulling-out of these yarns resulted in UD structures having a higher failure strain percentage in comparison with that of the T180 structures upon tensile loading.

In terms of composites' tensile moduli (fibre bundle stiffness calculated using the rule-of-mixtures) for the lines below the knee points in the stress-strain curves, the sequence in ascending order was: Hopsack<UD<T180<GVT. Out-of-plane undulation of the fibres was attributed as the main liable factor for the difference in modulus values. To analyse the results, micro-scale geometry was thoroughly studied in this research. Tapes contained more individualised elementary fibres; this ensured a better homogeneity of fibre distribution, consequently a reduced incidence of resin rich areas. The higher degree of individualised elementary fibres also ensured improved lumen impregnation. Also a greater number of elementary fibres being surrounded by matrix ensured a better stress transfer during tensile loading of the composites. Moreover, tapes contained reduced numbers of kink bands. During the drawing of the slivers, the fibres that were used to manufacture tape experienced strain hardening, which also improved the tensile properties of the fibres in comparison with the fibres in the Hopsack and UD structures. GVT showed enhanced stiffness properties compared with T180 due to the hybridisation effect induced by the glass veils. The sequence for the composites' moduli for the lines above the knee-points of the stress-strain curves was found to be the same as the moduli below the knee-points. Also the differences in percentage of the moduli values between the lines above and below the knee-points were found to be similar for all the structures. This inferred that the drop of a modulus upon tensile loading for FFRC above the knee-point was not a function of the fibre undulation, rather related to the fibre's morphology.

The strength values of the composites also followed the same sequence as that of the moduli values. Between UD and Hopsack, the latter showed reduced strength due to having more fibre waviness and kink-bands. In between UD and T180, the latter showed higher strength values. The reasons for superior strength values of T180 were attributed to the lesser out-of-plane waviness, higher extent of fibre individualisation, more homogeneity in fibre distribution, and greater extent of lumen impregnation. Both GVT and T180 exhibited approximately equal strength values. It was due to the fact that the hybridisation effect of GVT (inclusion of glass-fibre veil should have aided to attain a higher strength value) was mitigated by the random orientation of the glass-fibres (random orientation of fibres is a detrimental factor for tensile strength) in the surface veils of GVT.

As a further study, strain hardening of flax fibres during drawing should be thoroughly examined. Also, a detailed quantitative study to analyse the mutually dependant relationships among the elementary fibre individualisation, lumen impregnation, and their effects on the longitudinal tensile properties of the corresponding composites will be an important step to a better understanding of the load-bearing performances of FFRC.

## Acknowledgement

The main author would like to acknowledge the kind support ICT Ministry of Bangladesh for funding this research, National Composites Certification and Evaluation Facility for granting the access to their research facilities, and the School of Materials of The University of Manchester to provide the necessary support to accomplish this research.

## Funding statement

The PhD research was kindly funded by the Bangabandhu Fellowship project under the Ministry of Science and Information Technology of The People's Republic of Bangladesh.